# Smart ROI Detection for Alzheimer's Disease prediction using explainable AI

Atefe Aghaei[1], Mohsen Ebrahimi Moghaddam[1]

[1] Faculty of Computer science and Engineering, Shahid Beheshti University, Tehran, Iran

Corresponding Author: Mohsen Ebrahimi Moghaddam

Email: m_moghadam@sbu.ac.ir

ORCID iD: 0000-0002-7391-508X

**Abstract**
**Purpose**
**Predicting the progression of MCI to Alzheimer's disease is an important step in reducing the progression of the disease. Therefore, many methods have been introduced for this task based on deep learning. Among these approaches, the methods based on ROIs are in a good position in terms of accuracy and complexity. In these techniques, some specific parts of the brain are extracted as ROI manually for all of the patients. Extracting ROI manually is time-consuming and its results depend on human expertness and precision.**
**Method**
**To overcome these limitations, we propose a novel smart method for detecting ROIs automatically based on Explainable AI using Grad-Cam and a 3DCNN model that extracts ROIs per patient. After extracting the ROIs automatically, Alzheimer's disease is predicted using extracted ROI-based 3D CNN.**
**Results**
**We implement our method on 176 MCI patients of the famous ADNI dataset and obtain remarkable results compared to the state-of-the-art methods. The accuracy acquired using 5-fold cross-validation is 98.6 and the AUC is 1. We also compare the results of the ROI-based method with the whole brain-based method. The results show that the performance is impressively increased.**
**Conclusion**
**The experimental results show that the proposed smart ROI extraction, which extracts the ROIs automatically, performs well for Alzheimer's Disease prediction. The proposed method can also be used for Alzheimer's disease classification and diagnosis.**

## 1. Introduction

Today, in developed countries, the population is aging, and even if this fact is positive, it brings unwanted consequences such as an increase in various diseases such as dementia [1]. Alzheimer's Disease (AD), the most common form of dementia, is a major health care challenge in the 21st century. Alzheimer's disease is the sixth leading cause of death in the United States [2]. This disease is a neurodegenerative brain disorder with reduced cognitive function and there is no cure for it. Therefore, the mortality rate caused by Alzheimer's have increased by 68%, and life expectancy for Alzheimer's patients is now less than 7 years [3].

As we mentioned before, there is no certain cure for this disease [4], [5] only some definitive treatments are available which are effective for limited periods in subgroups of patients. Scientists believe that the most effective way to control the progression to AD is early diagnosis and an appropriate management strategy from the very beginning of cognitive decline. Therefore, many efforts have been made to find strategies for early diagnosis, especially in the early stages before the symptoms of the disease appear, in order to slow down or prevent the progress of the disease [6]. Since according to researchers, the brain changes caused by Alzheimer's disease can be seen up to 20 years before the disease manifests [7], the disease can be diagnosed by analyzing signals and images taken from the brain, including Magnetic Resonance Imaging (MRI), Positron Emission Tomography (PET), Electroencephalogram (EEG), and functional Magnetic Resonance Imaging (fMRI) years before AD appeared. Alzheimer's disease has three stages: Cognitively

Normal, Mild Cognitive Impairment (MCI), and Alzheimer's Dementia. MCI is a very critical stage for the diagnosis of Alzheimer's disease, which includes two outcomes: stable MCI (sMCI) are those people who do not develop Alzheimer's disease in the future, and progressive MCI (pMCI) are those who will develop AD in the future. According to the [8], 32% of people with MCI develop Alzheimer's dementia within five years.

Regarding to the importance of AD prediction, AI techniques have been considered in recent years by researchers and in this way, a lot of researches have been developed which use deep learning methods, including CNN, from MRI images. In some studies, whole 3D MRI images have been used to AD detection or prediction [9]–[11], in some other researches two-dimensional MRI images have been used to overcome complexity and data leakage of 3D subject-level studies[12]–[14]. Also, pre-trained convolutional neural networks including Res-Net designed for 2D natural images has been applied to 2D medical images with transfer learning [15]. However, in such approaches, converting the 3D image into 2D slices causes some useful information in the 3D image is lost. To overcome this problem, some studies focus on 3D patch-based classification. In these frameworks, the input consists of a set of 3D patches extracted from an image [16]–[18]. Most of these slices created from slicing 3D images do not have useful information because they may contain parts of the brain that are not affected by the disease. Therefore, methods based on regions of interest (ROI) focusing on regions known to be informative are used in many studies [19]–[21].

The main challenge of the region of interest-based methods is that only one (or more) region, which is also considered hippocampus region in most of the papers, is extracted while AD changes are widespread in several brain regions [15]. Also, selecting multiple regions at the same time may increase the complexity. To overcome this problem, in this article, a new method for smart ROI detection related to Alzheimer's prediction using the analysis of the whole brain 3D image of all patients is proposed. Although in the proposed method, ROI is detected per patient, it should be noticed that this region is not a different region for all subjects, it means that the region which is detected for each person may be common in most of the patients. Therefore, the main contributions of the paper are listed as follows:

- To overcome the problems of selecting manual ROI, an automatic region of interest detection method based on explainable AI for Alzheimer's prediction from 3D MRI images is proposed.
- We propose a method to predict Alzheimer's disease using detected ROI for each patient as input data.
- Also, according to our ROI detection method, to obtain an appropriate ROI for each subject, we propose a 3D classification model based on 3D convolutional neural network.

The rest of the paper has been organized as follows: In section 2, a history of similar studies is presented. In section 3, the proposed method is described in detail. Section 4 contains the experimental results. Section 5 presents a discussion about the paper.

## 2. Related work

In this section, we introduce some previous studies on Alzheimer's disease early detection using whole brain or extracting patches. Then we review some ROI based AD prediction.

### 2.1. Alzheimer's disease prediction

In Alzheimer's disease, symptoms usually appear after the age of 60, however, some forms of the disease develop very early in people with the gene mutation, even in their 30s to 50s [4].Alzheimer's disease causes structural and functional changes in the brain. As mentioned in the previous chapters, in Alzheimer's patients, there are several years between a healthy state and Alzheimer's. In the early stages of the disease, patients do not have obvious cognitively decline and after a while, they develop mild cognitive impairment (MCI) and gradually develop Alzheimer's. However, not all MCI patients convert to AD. Therefore, a major focus of current researches is on predicting the progression of MCI to AD. MRI is a neuroimaging technique that is commonly used to analyze and measure the structure of the brain and the brain changes. Recently, many studies have been conducted on the prediction of Alzheimer's disease from brain MIR images using deep learning methods. For example, in [22] a 3D CNN is proposed to predict rate of cognitive decline. Since using 3DCNN require a lot of data to train, some papers use transfer learning. For instance, in [23] a pre-trained 3D convolutional neural network is used to predict AD. In this study, also, age adjusted neural network is proposed. Researchers in another study proposed a modified version of ResNet which is ResNet_3D to predict progression of AD [24]. Also, in [25] VoxCNN (3D version of VGG) and ResNet_3D is applied to predict conversion of MCI to AD. Although studies result show customized CNN performs better if there is enough data, transfer learning obtains good results.

Sometimes to make the results more accurate, a combination of traditional methods and deep learning is used for AD prediction. For example, in [14] and [26] a combination of CNN and ensemble learning is proposed on whole brain MRI images and results show that ensemble learning enrich the accuracy. Generally, using 3D whole brain increases computational complexity. Therefore, some studies divide the whole brain image into some patches. A patch-based AD prediction model is proposed in [27]. Longitudinal data consists of three sets of images is used in this study and a dictionary of 2D patches from one 2D slice is made. Another patch-based AD prediction has been presented in [28]. In this paper, left hippocampus is extracted and local patches from left hippocampus assembled into 2.5 dimensions and fed into 2.5DCNN. In [29] the authors partitioned whole brain MRI images into some patches with the same size and they applied t test to sort these patches to obtain informative features. They proposed Patch-Net to extract features using 3D convolutional neural networks. As we mention in introduction, some patches may have no information for AD prediction, hence, more important patches should be selected. To this aim, in [30] patches based on AD related landmarks which are obtain by data-driven landmark discovery algorithm are extracted. Although this method solves the described challenge, it requires some pre-processing, including registration in the test phase, and therefore the accuracy of the model is limited to the accuracy of the registration method.

### *2.2. ROI-based Alzheimer's disease prediction*

Unlike whole brain-based studies which the input of the model is 2D or 3D brain MRI image, and patch-based studies which images are divided to some patches and mostly all of the patches are the input of model, some papers extract Region of Interest related to AD. Since gray matter is more effected by AD, in most of the studies whole brain is segmented to gray matter, white matter, and CSF, and the model is trained using gray matter only. For example, in [31], the gray matter of the brain is extracted and ensemble learning is used to classify the data and then for final prediction by voting, a group of deep belief networks is applied. To decrease computational complexity some researches extract smaller ROI instead of gray matter [32]. In [19], authors select hippocampus, amygdalae, and insularfrom axial sagittal and coronal slices as Three View Patches (TVP) ROI and fed these ROIs into CNN to feature extraction. In [33], using feature selection based on mutual information has been proposed in which five key features including left and right hippocampus, the thickness of the cortex of the quadruple bodies of the brain, the left upper temporal part and the right anterior part have been identified that have a positive effect on the classification. In this study, the selected features are classified using simple linear classification. As regard AD may affect different regions of the brain, selecting two or three ROIs may limit the performance hence in [34], 134 ROIs are selected and among them the most informative ROIs are identified such as caudal and rostral anterior cingulate gyrus, entorhinal, fusiform and insular cortex and the subcortical ROIs anterior corpus callosum and the left vessel, an ROI comprising lacunar alterations in inferior putamen and pallidum. In [35], two ROI based networks for converting NC to MCI are proposed. One network is single-ROI-based and the other is multiple-ROI-based. However, considering some ROIs increase computational complexity. Therefore, some papers select a small size of one ROI like hippocampus which is more important in Alzheimer's disease. For example, in [36] a transfer learning focusing on a few slices of hippocampus region is applied. To increase the accuracy, in [37], a recurrent neural network for time-based learning of longitudinal cognitive values of each subject and combining them with the hippocampus of the brain in the first examination have been developed to build a prognostic model of the progression of Alzheimer's disease. In [38], hippocampus and middle temporal gyrus are introduce as ROI. In this study, integrated regression framework combined with CNN is proposed.

One of the challenges of the ROI-based approaches which extract ROIs manually, is to select the best ROIs for every patient. To overcome this challenge and to establish a trade of between accuracy and time complexity, we introduce an ROI-based AD prediction approach which detect ROI automatically per person. Therefore, to decrease computational complexity, a few ROIs are used and since ROI is detected per person accuracy is not decreased.

### 3. **Proposed method**

In this section, proposed method is clarified. In section 3.1, an overview of the proposed method is presented. In section 3.2 image preprocessing is described. In section 3.3, the proposed method for extracting regions of interest based on interpretable AI and proposed 3DCNN for feature extraction is explained, and finally, in section

3.4, Alzheimer's prediction using the desired ROI is discussed.

### 3.1. An overview of the method

In fig.1 an overview of our proposed method is presented. As shown in the figure, in the proposed method, first, the images are preprocessed before entering to the model. The steps of the pre-processing are explained in section 3.2. After that, the 3D brain MRI images are fed into a 3D convolutional neural network classification, a binary classification to classify MRI images into s-MCI and p-MCI classes. Then, using the weights obtained from the last convolutional layer of the model, the more important parts of the images in decision making are extracted as Regions of Interest. The details of ROI detection are given in Section 3.3. After extracting the ROIs, these regions are fed into a 3D convolutional neural network and the final classification is obtained using these regions.

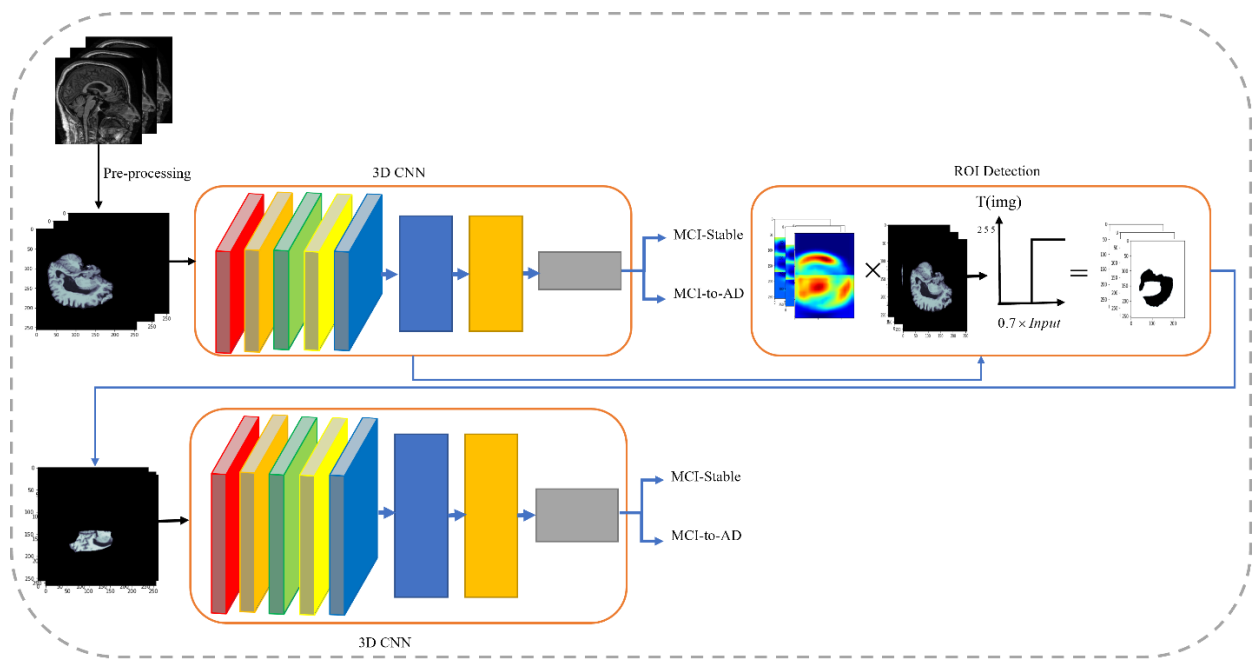

*Figure 1: An overview of the proposed method*

### 3.2. Pre-processing

Raw MRI images consist of skull and scalp. Therefore, at the first step of image pre-processing, brain is extracted from skull and scalp. Also, because of the imaging devices, MRI images may have artifacts like motion blur, intensity inhomogeneity, rotation or translation. In the proposed method, to artifact elimination, pre-processing, including B1 Correction, Grad Warp [39], and N3[40] has been applied to the images. Moreover, the size of the images is 256×256×256 so, in the last step, to reduce complexity of the model, the 32 slices from the middle of the image which include whole brain are sampled and the images have been resized into 128×128×32.

### 3.3. ROI Detection

In this section, ROI detection is described in details. The details of the method are shown in the figure 2. As shown in the figure, a 3D convolutional neural network which includes 3 CNN blocks is used. Each block consists of 3D convolution layers, Leaky ReLU activation function, Maxpooling, Batch Normalization and dropout. The number of 3D convolution layers of each block and the size of feature map in each step is written under the convolution layer in the figure 2. Also, the last layers of the 3D CNN are flattened layer, dense layer, Leaky ReLU activation function, BN, dropout and the last layer is softmax layer to classification. The implementation details are given in the figure description. First, the features of the images are weighted using the Grad-cam method [41] explained in section 3.3.1. Then a heatmap is generated for the pixels of the image based on the assigned weights. The heatmap, which has a value between zero and one (zero for lowest weight and one for highest weight), is multiplied to the image and a value is calculated for each pixel. A threshold is applied to pixel values and 30 percent of the more important pixels are obtained. The details of the ROI extraction using feature importance are described in section 3.3.2. finally, Alzheimer's disease is predicted using extracted ROIs. The details of the prediction are explained in section 3.4

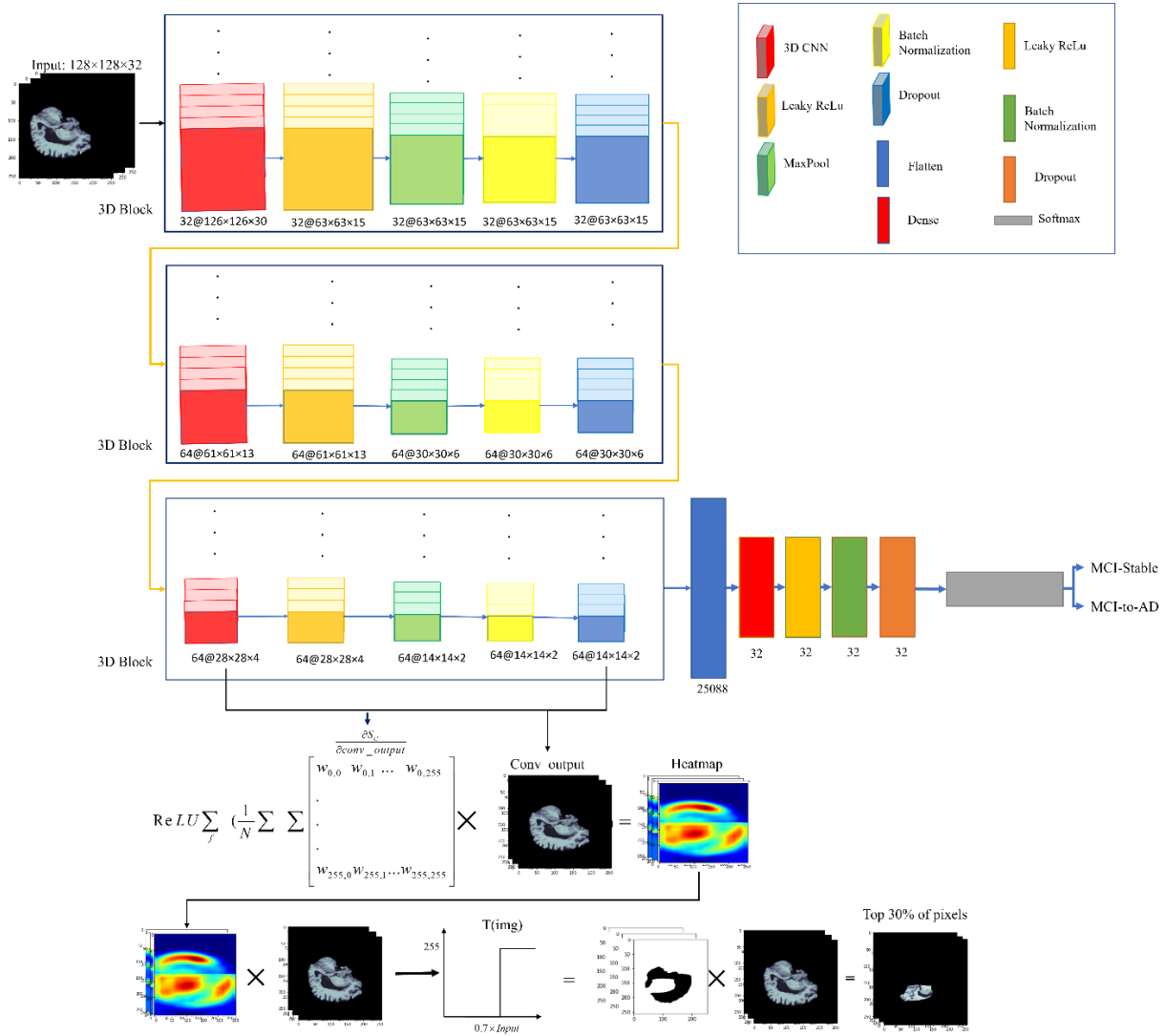

*Figure 2: the details of ROI detection. 3D images are fed into the 3D CNN which includes three 3D CNN blocks. The first block is designed using 32 3D convolutional layers with kernel size of 3×3×3, LeakyReLu activation function with α= 0.1, MaxPooling with size of 2×2×2 and dropout with rate= 0.3, the second and third blocks are designed using 64 3D convolutional layers with kernel size of 3×3×3, LeakyReLu activation function with α= 0.1, MaxPooling with size of 2×2×2 and dropout with rate= 0. the highest feature weights are extract from the last 3D block and ROIs are detected by multiplying the feature weights to the original image.*

### 3.3.1. Grad-Cam

Neurons of convolutional layers extract semantic information for each class in each image. Therefore, it is possible to understand which parts of the image had the most effect in decision making for that specific class according to this information. For example, to put a natural image into the dog class, we need to find information about the dog in the image, which the last convolution layer extracts from the image. Therefore, feature importance can be extracted from the last convolution layer of the model which is used for classification. In the proposed method, in order to obtain the importance of the features, Grad-CAM is used. First, the gradient of the last layer before softmax layer, $S_C$, with respect to each feature map of the last convolution layer, named $C_f$, is obtained, and then global average pooling is applied to the result using Equation 1 and the feature importance is calculated using Equation 2.

$$W_f = \sum_x \sum_y \frac{\partial S_c}{\partial C_f} \tag{1}$$

$$H_f = \text{Re} LU \sum_f (\frac{1}{N} W_f) \times C_f \tag{2}$$

According to the paper [41] to prove the eq.2, consider the Equation 3 to get the final score before softmax:

$$S_c = \sum_f W_f \frac{1}{Z} \sum_x \sum_y C_f \tag{3}$$

If the output of the global average poolnig is defined as follows:

$$G_f = \frac{1}{N} \sum_x \sum_y C_f \tag{4}$$

Therefore, equation 3 is rewritten as follows:

$$S_c = \sum_f W_f G_f \tag{5}$$

Therefore, according to equation 4 and 5, equations 6 and 7 are obtained:

$$\frac{\partial G_f}{\partial C_f} = \frac{1}{N} \tag{6}$$

$$\frac{\partial S_c}{G_f} = W_f = \frac{\partial S_c}{\partial C_f} N \tag{7}$$

Summing the values of both sides of equation 7, we get equation 8, and finally the equation 9 is reached. And to normalize the values, multiply it by 1/n, so we will reach the equation 1.

$$\sum_x \sum_y W_f = \sum_x \sum_y N \frac{\partial S_c}{C_f} \tag{8}$$

$$W_f = N \sum_x \sum_y \frac{\partial S_c}{\partial C_f} \tag{9}$$

The ReLU function is used to get the importance of the features that have a positive effect to decision making. In this way, the heatmap corresponding to the values of each of the feature maps, $H_f$, which shows the importance of the different parts of the image, is calculated using equation 2.

*3.3.2. ROI Extraction*

As shown in the figure 2, to extract the Regin of Interest, we resize the feature heatmap, $H_i$, and then multiply it to the original image, $x_i$, where *i* is the index of the image. After that, a threshold function is applied to the intensity values of the image and a 0-1 mask is, $M_i$, made for each image as follows:

$$M_i = \begin{cases} 255 & \text{if } x_i \times H_i \geq a \times x_i \\ 0 & \text{otherwize} \end{cases} \qquad (10)$$

In this equation, '$a$' is threshold value which means the intensity of the regions of original image where the heatmap is greater than $a$, convert to 255, and the intensity of other regions is zero and then we normalize the mask by dividing by 255. Therefore, a zero-one mask is created. The mask is made using equation 10 is multiplied to the original image and the ROI which we consider the top $(1-a) \times 100$ percentages of the important pixels of the image is extracted using equation 11:

$$x'_i = M_i \times x_i \qquad (11)$$

### 3.4. Alzheimer's Disease prediction

In the last step, to predict the Alzheimer's disease using MCI patient's 3D MRI images, the 3D convolutional model described in section 3.3 is used (Fig. 3). As shown in the figure, the model is the same as initial model which is used to extract Region of Interest. As mentioned in the section 3.3, the model is the two-class classification model with 3 blocks of 3D convolution layers and the last layers are flatten, dense, and softmax layers. Also, the output of the model is two classes include MCI-Stable and MCI-to-AD. To this purpose, ROIs extracted in section 3.3 are fed into the model instead of the original images.

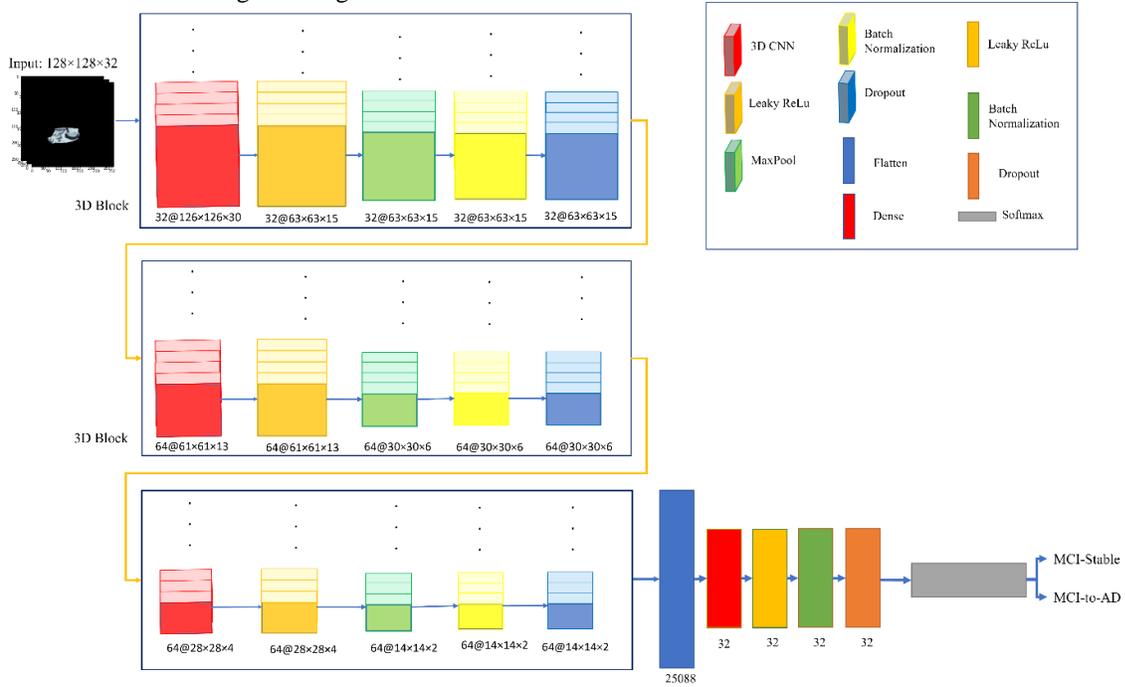

*Figure 3: the last classification to make final decision using ROIs as input. The model is 3DCnn includes 3 CNN blocks which is described in fig. 2.*

## 4. Experimental results

In this section the dataset and the results of the proposed method are discussed.

### 4.1. Dataset

Some T1 weighted 3D MRI images from famous and free ADNI (Alzheimer's Disease Neuroimaging Initiative) [42] dataset which consists of about 85 thousand of subjects are selected for this paper. ADNI is a multicenter study designed for the early detection and tracking of Alzheimer's disease and contains clinical, imaging, genetic, and biochemical biomarkers. This study has been collecting data from volunteers since the early of 2004 in ADNI1, ADNI2, ADNI3 and ADNIGo phases and the subjects are classified into Cognitively Normal (CN), Mild Cognitive Impairment (MCI) and Alzheimer's disease (AD). In this paper, 176 MCI patients (88 MCI-Stable and 87 MCI-to-AD patients) are selected. The study is longitudinal study and five sets of the images every six months (from baseline to 24 months) are used. Tables 1 shows the details of the data.

*Table 1: Distribution of volunteers based on Age*

|  | category | %Patients |
|---|---|---|
| **Age** | 60-69 | 15.5% |
|  | 70-79 | 41% |
|  | 80-89 | 41% |
|  | 90-93 | 2.5% |
| **Gender** | Male | 67% |
|  | Female | 37% |

### 4.2. Implementation details

The proposed networks are implemented in the Keras library in python 3.9 using GPU Tesla T4 and Intel(R) Xeon(R) CPU @ 2.20GHz. The 3D CNN model contains three 3D CNN blocks and some dense layers. The first block is designed using 32 3D convolutional layers with kernel size of 3×3×3, LeakyReLu activation function with α= 0.1, MaxPooling with size of 2×2×2 and dropout with rate= 0.3, the second and third blocks are designed using 64 3D convolutional layers with kernel size of 3×3×3, LeakyReLu activation function with α= 0.1, MaxPooling with size of 2×2×2 and dropout with rate= 0.3, also the size of the output feature maps of each layer is written under each feature map in the figure 2. We use Adam optimization function using the dynamic learning rate with the initial learning rate of 0.01 and Categorical_Cross_Entropy loss function.

### 4.3. Analysis of the proposed method

In this section, the proposed method is evaluated on MCI conversion (MCI-Stable vs MCI-to-AD classes) using the introduced dataset. We use both hold out and k-fold cross validation to obtain certain results. Since there are about 5 sets of MRI images per subject, to obtain more accurate results, we hold out 10 percentages of the subjects as test data and then split training data into five folds to create validation and train dataset. The model is trained five times using each set of training and validation data. Finally, the experimental results are the average of 5 models results on test data. The evaluation criteria are Accuracy, Precision, Recall and F1-Score which are defined in eq 12 to equ 15 respectively. Also, Area Under the Recursive Operating characteristic Curve (ROC) is calculated using True Positive Rate (TPR) and False Positive Rate (FPR) to verify the model.

$$\text{Precision} = \frac{TP}{TP + FP} \quad (12)$$

$$\text{Recall} = \frac{TP}{TP + FN} \quad (13)$$

$$F1 = 2 \times \frac{\text{Precision} \times \text{Recall}}{\text{Precision} + \text{Recall}} \quad (14)$$

$$Accuracy = \frac{TPR + TNR}{2} = \frac{TP + TN}{TP + FN + TN + FP} \quad (15)$$

#### 4.3.1. The results on whole 3D brain MRI images

First, the proposed 3D convolutional method is evaluated on original whole brain 3D MRI images. Since, some slices do not have meaningful information and there is no brain in most of the slices, to decrease computational complexity and increase the accuracy, 70 slices from the middle of the MRI images, where there is whole brain, are selected and then the images are resized into 128×128×32. After that data augmentation is applied to the training set to increase the amount of data. Since every pixel of the brain image has important information, flipping and rotation have been chosen as augmentation technique. Therefore, the images are vertically and horizontally flipped. Also, to rotate the images, a vector of some rotation angles such as, -10, -5, 5, 10 are defined and each time one of these angles are randomly selected. Both original and created images are fed into the proposed 3D CNN and the model is trained using these images. Finally, test set are fed into the trained model to validate the generalization of the model. The confusion matrix and the ROC curve of the average results on the test data are illustrated in fig.4. Also, the average of accuracy, Precision, Recall and F1-Score using 5 models on test data is shown in table 2.

Table 2: The average of accuracy, Precision, Recall and F1-Score using 5 folds on original whole brain images

| Accuracy | Precision | Recall | F1-score |
|---|---|---|---|
| 0.80 | 0.78 | 0.84 | 0.81 |

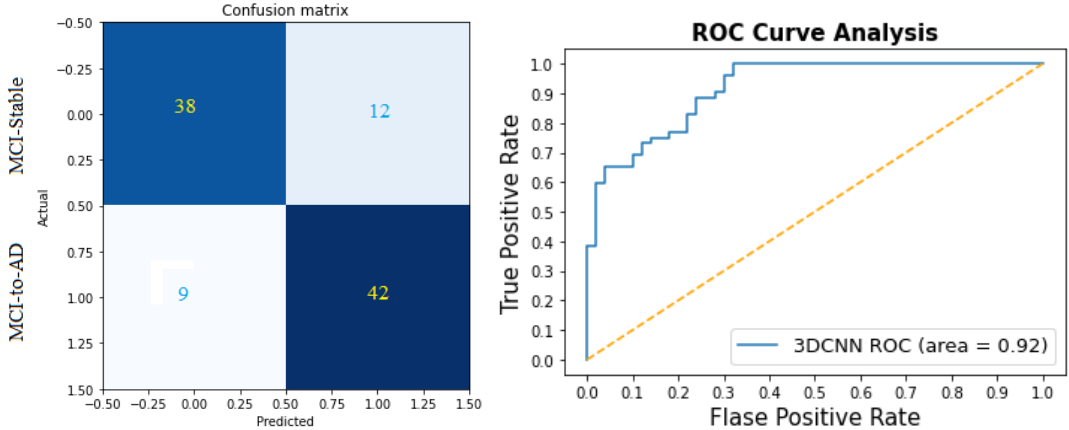

Figure 4: confusion matrix and ROC curve of the proposed model on the average of 5 folds.

### 4.3.2. Extracting ROI

As mentioned in the introduction, in order to avoid time complexity and sometimes even to increase accuracy, instead of using the whole brain image, ROI is extracted and classification is performed only according to those parts. One of the challenges in this field is that only one region, hippocampus region in most of the papers, is extracted while AD changes are widespread in several brain regions specially in early stage. To solve this challenge, in this article, instead of extracting ROI manually, ROIs are detected automatically per patient. In this section, the results of ROI detection are explained. As mentioned in section 3.3.2, to extract automatic ROI, after obtaining the heatmap of the feature map's weights, a zero-one mask is created for every image where top $(1-a)\times 100$ percentages of the feature map's weights are considered as one and the other regions are zero. In the figure 5 the heatmap of some images using this value is shown. The heatmap color is from dark blue to dark orange and the features with greater weights are dark orange, which means, the features in dark orange color are the most important features.

According to the equation 10, there is a threshold value, $a$, which is a number between zero and one that decide what percentage of the regions of the original image should be considered as Region of Interest. In other words, this number is a threshold selects the values of the image where heatmap of their feature weights is greater than this number, then these pixels are selected as ROI. Since this value is a hyperparameter, the appropriate measure should be chosen for it. To this end, some values are tested for this hyperparameter and for every one of them all of the introduced criteria is calculated using 5-fold cross validation.

The ROIs extracted from the images are fed into the classification model and the results are reported. Training and test sets of the whole brain images have been saved for this section. The ROIs of all of the samples of the training sets (there are 5 training and 5 validation sets) which are classified correctly by the model is extracted and then these ROIs fed into 3D model as training set and the model is trained again using ROIs only. After that, ROIs of the held-out test data are extracted and the trained model is tested using these ROIs. 5 models using 5-fold training and validation data are obtained. The experimental results are the average of 5 models obtained from 5 folds on ROIs which are extracted from test data. The evaluation criteria are Accuracy, Precision, Recall and F1-Score which are defined in eq. 12-15 respectively. The average of the results on 5 folds are shown in the table 3.

As shown in Table 3, second to sixth rows, to find the optimum percentage of the super-pixels as ROI, we consider 0.5, 0.6, 0.7, 0.8, 0.9 for '$a$'. According to the results, it can be seen that the performance of the model is enhanced with the increasing amount of '$a$'. However, the criteria remain fixed at values greater than 0.7, hence, any value in the range of 0.7 to 0.9 can be selected. Since this paper aims to find important regions involved in Alzheimer's disease in different patients, we have chosen the greatest value to show that the model extracts all important regions.

To prove the reliability of the results, we once consider the heatmap lower than 0.3 as ROI instead of higher than 0.7. In this case, since high weights have the greatest impact on obtaining results, low accuracy should be obtained by considering low weights. The last row of the table 3 shows the results obtained from this assumption. According to the result, the accuracy of the model on test data using 30 percentages of pixels with higher weights about 98% whereas, the accuracy is 50% using 30 percentages of pixels with lower weights which means the model obtain the lowest performance. Therefore, we can claim that our approach to extract smart ROI performs well.

*Table 3: the results obtained using different values for $a$*

| Train-Test (%) | $a$ | Percentage of feature map's weight | Accuracy | Precision | Recall | F1-score |
|---|---|---|---|---|---|---|
| 80-20 | 0.9 | **Top 10%** | 0.98 | 0.98 | 0.99 | 0.98 |
| 80-20 | 0.8 | **Top 20%** | 0.98 | 0.99 | 0.98 | 0.98 |
| 80-20 | 0.7 | **Top 30%** | 0.98 | 0.98 | 0.99 | 0.98 |
| 80-20 | 0.6 | **Top 40%** | 0.93 | 0.88 | 0.95 | 0.91 |
| 80-20 | 0.5 | **Top 50%** | 0.90 | 0.84 | 1.0 | 0.91 |
| 80-20 | 0.7 | **Low 30%** | 0.50 | 0 | 0 | 0 |

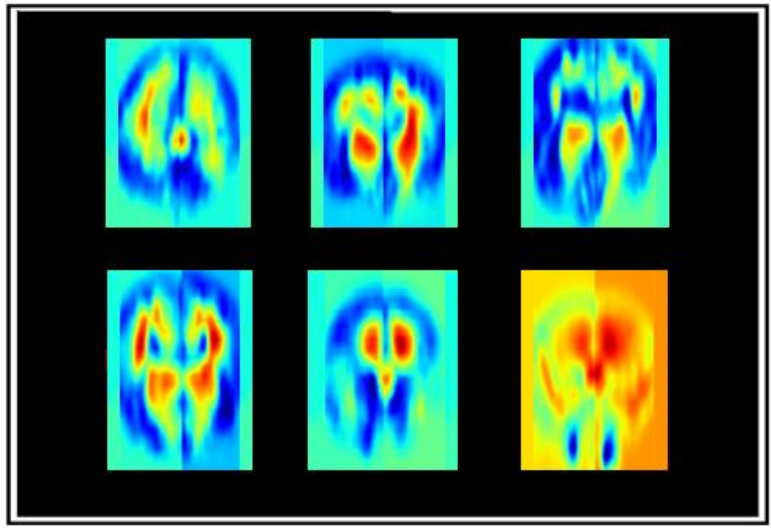

*Figure 5: the heatmap of some images using 0.7 for '$a$' value. In this figure the feature importance is shown by a value between zero and 1. The dark orange are the features with the highest weights.*

The results of these criteria using the selected value for $a$ (0.7) per each fold have been shown in table 4. Also, the average of Area Under the Recursive Operating characteristic Curve (ROC) and confusion matrix using $a$ equal to 0.7 is shown in the fig. 7.

According to the table 4, the average of the accuracy, Precision, Recall, and f1-score on the best extracted ROIs are 98.6%, 98%, 99.2% and 98.6% respectively and compared to the table 2 all of the results are increased when the best ROIs are used instead of whole brain images. Also, the AUC is 1 using ROIs which increased 0.08 compared to the results on whole brain images.

Table 4: The average of Accuracy, Precision, Recall and F1-Score using 5 models on extracted ROIs of test data.

| Fold | Accuracy | Precision | Recall | F1-score |
| --- | --- | --- | --- | --- |
| Fold-1 | 0.99 | 0.98 | 1 | 0.99 |
| Fold2 | 0.98 | 0.96 | 1 | 0.98 |
| Fold3 | 0.98 | 0.98 | 0.98 | 0.98 |
| Fold4 | 0.98 | 0.98 | 0.98 | 0.98 |
| Fold5 | 1 | 1 | 1 | 1 |
| Average | 0.986 | 0.98 | 0.992 | 0.986 |

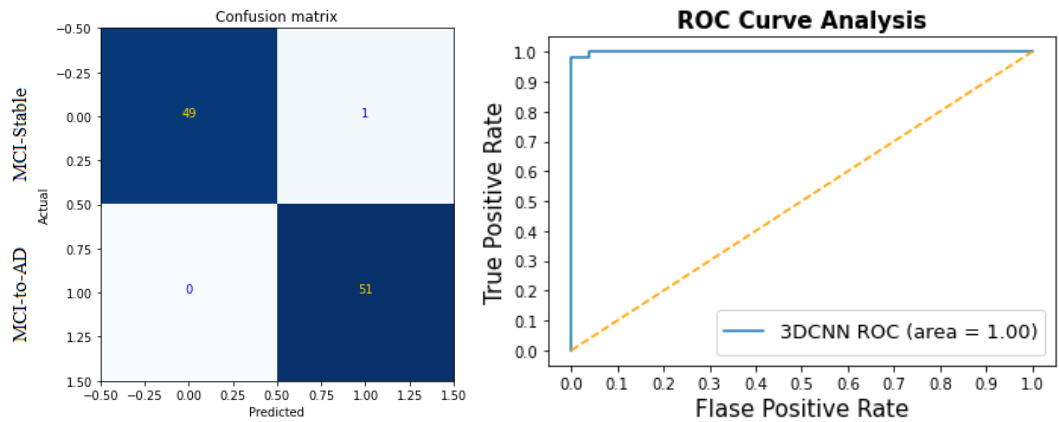

Figure 6: The average of confusion matrix and ROC curve using 5 folds on extracted ROIs of test data.

### 4.3.3. Representing the extracted ROI

According to the previous section, the best value for '$a$' is 0.7. After creating the mask using '$a$' based on equation 10, the ROI is obtained using the created mask and equation 11. In this section, extracted ROI for images are analyzed. Some of extracted ROIs are shown in the fig. 6. To show the extracted ROI, 3D ROIs are converted to 2D and one of the slices from the middle of the 3D image is chosen. Based on [43], the most important changes occurred in the MCI stage are shrinkage of hippocampus, cortical thinning, and enlargement of ventricles filled with cerebrospinal fluid. Also, due to the recent studies, contrary to traditional belief, in very early stage of Alzheimer's disease, entorhinal cortex which is the location of tau protein is affected before hippocampus changes [44]. Therefore, according to the fig.6, these ROIs of the six samples are marked. As can be seen, in every patent, some of the introduced ROIs are extracted. Also, other parts of the images are also extracted as ROI, which are not shown in the figure since they do not have many repetitions in other images.

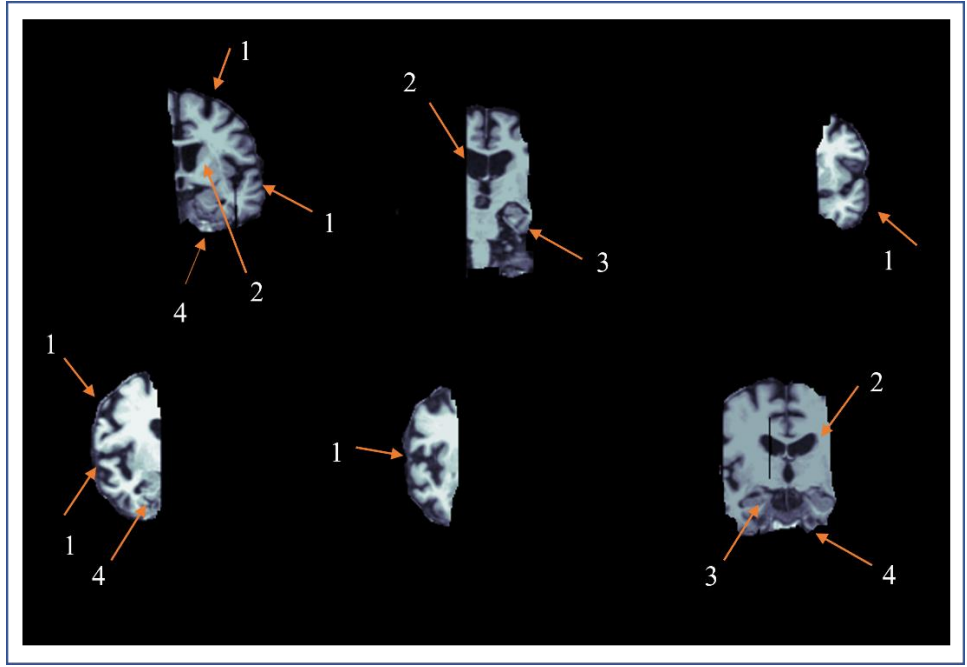

*Figure 7:the results of six sample of the test data. Results show four different ROIs which are more important for AD prediction include 1. cortical area 2. ventricles filled with cerebrospinal fluid 3. hippocampus, and 4. entorhinal cortex.*

To analyze the obtained results, the percentage of repetition of each of the important sections that are affected by Alzheimer's disease based on the articles (which is mentioned in previous paragraph) is shown in the table 5. According to the table, the sum of the percentages is not equal to 100, because as seen in the figure, more than one ROI is extracted in some images. Due to the table, the more repetition is for cortical thinning because shrinkage of the cortical is occurred in the whole gray matter and therefore, this ROI is extracted for the most of the samples. After that, hippocampus and entorhinal cortex has the most repetition in the samples. According to the researches these two parts are the most important parts in the early stage of the Alzheimer's disease. Since the images are not from one time, it means that the images are from six months before AD, 12 months before AD, 18 months before AD and 24 months before AD, the parts are affected during the time are a little changed. For example, shrinkage of the hippocampus is more when the patients get AD than before 24 months and in 24 months before AD entorhinal cortex is more effected. Therefore, the repetition of these two ROIs is almost the same. The last ROIs with high repetition are ventricles filled with cerebrospinal fluid. This ROIs which are more important in the final stage, are repeated in 40% of the samples.

*Table 5: the percentage of repetition of each of the proposed ROIs sections shown in the fig.6 that are affected more by Alzheimer's disease based on [43],[44].*

| Correlation | 1 | 2 | 3 | 4 |
|---|---|---|---|---|
| %Test samples | 70% | 40% | 60% | 60% |

### 4.4. Comparison with the other models

In this section, a comparison between the proposed model and some state-of-the-art models are discussed. In table 6 some recent studies on AD prediction are introduced. All of the compared studies are AD prediction using structural MRI images of ADNI dataset to make comparison more acceptable. The state-of-the-art studies are some of the patch-based and ROI-based studies which are introduced in related works section. As shown in the table 6, three patch-based studies and ROI-based studies are selected. In ROI-based studies the most effective parts of the brain to AD based on the literature are extracted as ROI. For example, in [45] hippocampus, in [46] the hippocampus, fusiform, and inferior temporal gyrus and in [33], key features including left and right hippocampus, the thickness of the cortex of the quadruple bodies of the brain, the left upper temporal part and the right anterior part have been identified that have positive effect on the classification. In [18], 3D patches are extracted and to increase the accuracy a dual multi-instance attention based deep neural network is proposed. The results of this paper are better than ROI-based studies but as we mentioned before, some of these patches may don't have important information for this reason the accuracy may decrease. To solve this problem, in [28] left hippocampus is extracted from the image and local 2.5D patches are extracted from only this ROI. In ROI-based studies some ROIs are extracted manually and prediction is done using those ROIs. Authors in [30] believe that manually ROI extraction limit the performance of the prediction because of 1) defining the specific ROIs and 2) extracting effective disease-related features. To solve this problem, they propose a landmark-based framework which extract informative patches using AD-related landmarks. The results of this paper show they achieve good results. The aim of our proposed method, which extracts automatic ROI, is the same and it gets the best results compare to the other studies. However, our proposed feature extraction model, even based on whole brain, achieve good results which is obvious in the table.

*Table 4: the comparison of the proposed model and some state-of-the-art methods on Accuracy, Sensitivity, and Specificity*

| Method | Number of patients | Whole brain | Patch-based | ROI | Accuracy (%) | Sensitivity (%) | Specificity (%) |
|---|---|---|---|---|---|---|---|
| Patch_base_2.5D CNN (2018) [28] | 264 | | ✓ | | 79.9 | 89.6 | 68 |
| LDA (2015) [33] | 388 | | | ✓ | 71.9 | 69.6 | 73.6 |
| Attention+MIL + CNN (2020) [18] | 289 | | ✓ | | 80.2 | 77.1 | 82.6 |
| LDMIL (2018) [30] | 164 | | ✓ | | 78.3 | 47.3 | 83.2 |
| DCNN (2022) [46] | 115 | | | ✓ | 74 | 62 | 78 |
| DCNN (2022) [45] | 381 | | | ✓ | 75.85 | 0.66 | 0.8 |
| 3D-CNN (Proposed) | 175 | ✓ | | | 80 | 0.84 | 0.78 |
| 3D-CNN (Proposed) | 175 | | | ✓ | 98.6 | 99.2 | 98 |

5.  Discussion
    .
    This section is divided to three parts. In the first part the importance of Alzheimer's disease prediction and our dataset is discussed, in the second part our proposed method and the automatic ROI extraction is explained, and the last part is about limitations of the paper and our future works.

*5.1. Alzheimer's disease prediction*

According to the literature, Alzheimer's disease, the most common form of dementia, is a neurodegenerative brain disorder with cognitive decline. Also, there is no cure for this disease, therefore the most effective way to control the progress of this disease is early detection. This paper proposed an approach to early detection of AD using T1 weighted MRI images of 176 MCI patients (including sMCI and pMCI) selected from ADNI dataset. The data are longitudinal including almost five sets of images per patient (some of the patients has less than five sets). First images are preprocessed, brain is extracted from the images, after that, since deep learning methods get remarkable results in AD prediction, we proposed a 3D CNN to extract the features. The results show our model obtain good results compare to other methods.

*5.2.   Automatic ROI detection*

Region of interest is a part of the image which is more important for the specific task and in some studies ROI is extracted from the image and classification is perform using those parts. As explained in introduction and result section, in some state-of-the-art researches one or some ROIs are extracted for AD prediction and these ROIs mostly include hippocampus, cortex thickness, temporal gyrus and two or three other parts. Results show extracting some specific regions and omitting the rest of the brain affect the results. To overcome this issue, in some studies patches are extracted from the image instead of ROI but most of the patches do not have useful information.  As discussed in the previous section some studies combine patch-based and ROI-based approach. In this paper we propose an automatic ROI detection approach which extract ROIs for each patient instead of considering some specific ROIs for all of the patients. In the proposed method, first whole 3D brain images fed into 3DCNN model, after that, the most important regions of the image using the highest feature weights are extracted as ROI based on decision making. We consider a threshold to obtain the highest weights and according to the results the best threshold is 0.7. Finally, the extracted ROIs are fed into 3DCNN classifier model to make the final decision. According to the results, the accuracy of the proposed model using the extracted ROI is 98.6% and 1 respectively, whereas, the accuracy and the AUC of the proposed model using whole brain image is 80% and 0.92. The results show the accuracy increased 18.6% using ROIs instead of whole brain images, Also the AUC is improved 0.08 using the extracted ROI

*5.3.   Limitations and future work*

Although the proposed method obtains good results, there is some limitation in the approach. The first limitation is using MCI patients who are convert (or not convert) to AD after 24 months but the most important issue is predicting AD in early stages for example 10 years earlier than conversion. Therefore, in the future work we will consider the normal subjects which will convert to AD after years. The second limitation is that in our proposed method MRI images are used only while metadata like age, gender, education and so on are important biomarkers. Hence, in the future we will use these biomarkers as well as MRI images.